# All-fiber upconversion high spectral resolution wind lidar using a Fabry-Perot interferometer

Mingjia Shangguan[1,2,6], Haiyun Xia[1,3,6], Chong Wang[1], Jiawei Qiu[1], Guoliang Shentu[2,4], Qiang Zhang[2,4,5], Xiankang Dou[1,*], Jian-wei Pan[2,4]

[1]CAS Key Laboratory of Geospace Environment, USTC, Hefei, 230026, China

[2]Shanghai Branch, National Laboratory for Physical Sciences at Microscale and Department of Modern Physics, USTC, Shanghai, 201315, China

[3]Collaborative Innovation Center of Astronautical Science and Technology, HIT, Harbin 150001, China

[4]Synergetic Innovation Center of Quantum Information and Quantum Physics, USTC, Hefei 230026, China

[5]Jinan Institute of Quantum Technology, Jinan, Shandong 250101, China

[6]These authors contributed equally to this work

*dou@ustc.edu.cn

## Abstract

An all-fiber, micro-pulse and eye-safe high spectral resolution wind lidar (HSRWL) at 1.5 μm is proposed and demonstrated by using a pair of upconversion single-photon detectors and a fiber Fabry-Perot scanning interferometer (FFP-SI). In order to improve the optical detection efficiency, both the transmission spectrum and the reflection spectrum of the FFP-SI are used for spectral analyses of the aerosol backscatter and the reference laser pulse. The reference signal is tapped from the outgoing laser and served as a zero velocity indicator. The Doppler shift is retrieved from a frequency response function Q, which is defined as the ratio of difference of the transmitted signal and the reflected signal to their sum. Taking advantages of high signal-to-noise ratio of the detectors and high spectral resolution of the FFP-SI, the Q spectra of the aerosol backscatter are reconstructed along the line-of-sight (LOS) of the telescope. By applying a least squares fit procedure to the measured Q spectra, the center frequencies and the bandwidths are obtained simultaneously. And then the Doppler shifts are determined relative to the center frequency of the reference signal. To eliminate the influence of temperature fluctuations on the FFP-SI, the FFP-SI is cased in a chamber with temperature stability of ±0.001during the measurement. Continuous LOS wind observations are carried out on two days at Hefei (31.843 ° N, 117.265 ° E), China. Horizontal detection range of 4 km is realized with the temporal resolution of 1 minute. The spatial resolution is switched from 30 m to 60 m at a distance of 1.8 km. In the meantime, LOS wind measurements from the HSRWL show good agreement with the results from an ultrasonic wind sensor (Vaisala windcap WMT52). Due to the computational expensive of the convolution operation of the Q function, an empirical method is adopted to evaluate the quality of the measurements. The standard deviation of the wind speed is 0.76 m/s at the 1.8 km. The standard deviation of the retrieved bandwidth variation is 2.07 MHz at the 1.8 km.

## 1. Introduction

Doppler wind lidars (DWL) have shown their efficacy in the remote measurement of spatially resolved atmospheric wind velocities in many applications, including measurement of atmospheric boundary layer[1], aircraft wake vortices[2-5], air turbulence [6, 7] and wind shear[8]. With high spectral resolution techniques, atmospheric temperature [9-14], wind [12, 15] and aerosol optical properties [10, 16] can be retrieved. The measurement principle of the DWL relies on the detection of the radial Doppler shift carried on light backscattered off the moving particles in atmosphere. The DWL is generally classified into two main categories: the coherent (heterodyne) detection lidar (CDL) and the direct detection lidar (DDL). In the CDL, the backscattered signal is optically mixed with a local oscillator, where the resulting beat-signal' frequency, except for a fixed offset, is the Doppler shift due to the moving particles. In the DDL, no local oscillator is required; instead an optical frequency discriminator or an optical spectrum analyzer converts the Doppler shift into an irradiance variation.

Although the sensitivity of heterodyne detection technique is high because the local oscillator amplifies the signal, optical systems used for coherent detection are generally more challenging [17]. Much of this is due to the need to match wave-fronts from the local oscillator and backscattered signal at the heterodyne mixer accurately. Moreover, in the CDL, the beating signal is recorded using an analog–to-digital converter (ADC). High sampling frequency is required to improve the spatial resolution, resulting heavy burden in time-consuming post-processing. In addition, the coherent system is strongly affected by shot noise, phase noise and the relative intensity noise of the local oscillator. Furthermore, the remote sensing range is limited by the laser coherence length [18]. Finally, In order to distinguish the sign of the Doppler shift, acousto-optic modulator must be adopted as the frequency shifter.

In comparison, the DDL has several advantages including having a large remote sensing range not limited by the laser coherence, inherent sign discrimination of Doppler shift, more robust to phase aberration, much simpler data acquisition and data processing. According to different implementations, the direct detection techniques fall into two types. The first one is edge technique, in which one or more narrowband filters are used and the Doppler frequency shift is determined from the variation of the transmitted signal strength through the filter. The other is called fringe-imaging technique, where the Doppler shift is determined from the radial angular distribution or spatial movement of the interference patterns through an interferometer. The two implementations have been compared theoretically [19].

In this work, a new high spectral resolution wind lidar (HSRWL) at 1.5 μm is proposed, where the aerosol backscatter spectra along the line-of-sight of the telescope are analyzed by using a fiber Fabry-Perot interferometer and detected with two upconversion detectors. By applying direct spectral analyses, the center frequencies and the bandwidths of these spectra are retrieved simultaneously. The Doppler shifts are determined by calculating the difference in center frequencies of the aerosol backscatter and the reference signal.

In many applications, eye safety, long-range sensing capability and large dynamic range of the wind speed measurement are the primary considerations in Doppler wind lidar design. Direct-detection Doppler lidars usually operate at wavelength of 1.064 μm [20], 0.532 μm [21] or 0.355 μm [22, 23], which produced by the fundamental, second or third harmonics of an Nd: YAG laser. The proposed lidar operates at a wavelength of 1.5 μm, a standard wavelength of telecommunications industry, providing many advantages. Firstly, the 1.5μm wavelength is at the edge of the near infra-red range, making it more conductive to be eye-safe. Secondly, optical fiber components and devices developed for optical communications are commercial available. Thirdly, components have been designed to be environmentally hardened, making the system highly reliable. In addition, the advantages of the longer wavelength include: more relaxed requirement on optical surface quality and decreased sensitivity to atmospheric refractive turbulence.

Measurement dynamic range of wind speed is a significant issue if an aerosol lidar system is required to operate at higher altitudes where atmospheric motions can be considerable. For the proposed HSRWL systems, the wind speed dynamic range is determined by the free spectral range (FSR) of the FFP-SI, and is typically larger than of the systems based on the edge technique, where the dynamic range is set by the etalon passband width. For heterodyne detection system, the dynamic range is limited by the bandwidth of the detectors and the ADC

(higher velocity measurements require higher bandwidth detectors and higher speed ADC in order to satisfy the Nyquist sampling criterion).

In comparison, the lidar system based on the fringe-imaging technique is found to have large measurement dynamic range, similar to the proposed HSRWL [19]. However, an intrinsic property of the fringe-imaging is the requirement of a multielement detector that is matched to the Fabry-Perot fringe pattern. This results in a detector package that can be more complicated than the singe-element detector used for the HSRWL.

In our previous work, temperature lidar [15] and Doppler wind lidars [20, 22] based on free-space Fabry-Perot interferometer and photon-counting detectors have been demonstrated. However, it is hard to eliminate the parallelism error of the reflecting mirrors during the cavity scanning and the mode-dependent spectral broadening due to its illuminating condition. In this work, this problem is solved by using a lensless FFP-SI.

In contrast to our prevision aerosol lidar at 1.5 um [24], the HSRWL adopts upconversion detectors using all-fiber configuration. Recently, an intracavity upconversion system has been demonstrated for detection of atmospheric $CO_2$ [25]. A Doppler velocimetry has been demonstrated before by using a scanning Fabry-Perot interferometer and an InGaAs photodetector [26, 27]. Although InGaAs detector is commercial available, its high after-pulse probability distorts the raw signal significantly and the time gating operation mode exhibits periodic blind zones [28]. Different from the reported system for hard target detection, the FFP-SI and upconversion detectors are adopted for wind sensing. Both the transmission spectrum and the reflection spectrum of the FFP-SI are applied to analysis the spectra of the aerosol backscatter.

## 2. Principle

The key instrument inside the optical receiver of the HSRWL is the FFP-SI. The cavity of the FFP-SI is formed by two highly reflective multilayer mirrors that are deposited directly onto two carefully aligned optical fiber ends [29]. The anti-reflection coated fiber inserted in the cavity provides appropriate confined light-guiding and eliminates secondary cavity. Due to the fiber-based design, the fiber-to-fiber insertion losses are low. Here, a stacked piezoelectric transducer (PZT) is used to axially strain a short piece of single-mode fiber that inserted in the cavity. Frequency scanning of the FFP-SI is achieved by scanning the cavity length [20]. A step change of the cavity $\Delta l$ is related to the frequency sampling interval $\Delta v$ as

$$\Delta v/v_0 = -\Delta l/l, \qquad (1)$$

where, $l = 25.59$ mm is the cavity spacing, $v_0$ is the frequency of the incident laser. As an example, if one needs to increase the frequency of FFP-SI over 4.02 GHz, the cavity spacing should be shrunk 532 nm. Note that, after the spacing change, the free spectral range (FSR) of the FFP-SI also increases 0.08 MHz, which is ignorable in this work.

The HSRWL using the FFP-SI is proposed, in which both the transmission spectrum and the reflection spectrum of the FFP-SI are used for spectral analyses. As shown in Fig. 1(a), in a calibration experiment, a continuous-wave (CW) monochromatic laser is used. By changing the voltage fed to the PZT, the transmission spectra and the reflection spectra are recorded simultaneously using an oscilloscope. In the frequency domain, the transmission and reflection of a Fabry-Perot interferometer is periodic with a constant free spectral range (FSR). In Fig. 1, a time to frequency converter is performed by using a constant FSR of 4.02GHz. If the FFP-SI analyzes both aerosol backscatter and the reference laser, the Doppler shift will shift the spectra of aerosol backscatter to one side of the reference peak. The red-shifted or blue-shifted determine the sign of the Doppler shift, namely the sign of velocity. Then, the possible dynamic range for velocity measurement with sign-discrimination spans from -1/2 FSR to 1/2 FSR. Assuming the FSR of the FFP-SI is 4.02 GHz, the maximum wind speed dynamic range of the proposed HSRWL is from -1557 m/s to -1557 m/s for the wavelength of 1550 nm.

Note that, given a constant scanning step, the wider the scanning range, the longer time it takes. Thus, for moderate wind speed measurement range, such as in atmospheric boundary layer application, it is more practical to select a moderate frequency scanning range in order to increase the temporal resolution.

In order to retrieve the Doppler shift, spectral analyses is performed as follows. For the pulsed laser, the measured transmission spectrum of the aerosol backscatter $T(v,v_c,v_D,\Delta v_M)$ is the convolution of the spectrum of aerosol backscatter $I(v,v_c,v_D,\Delta v_M)$ and the transmission function of the FFP-SI $h(v)$:

$$T(v,v_c,v_D,\Delta v_M)=h(v)*I(v,v_c,v_D,\Delta v_M), \qquad (2)$$

where $*$ denotes the convolution, $v$ is the optical frequency relative to the center of the spectrum of the reference signal, $v_c$ is the center frequency of the spectrum of the reference signal, $v_D$ is the Doppler shift carried on atmospheric backscattering, $\Delta v_M$ is the half-width at the 1/e intensity level of the spectrum of aerosol backscatter.

The FFP-SI is fabricated with single-mode fiber with a negligible divergence in the cavity, thus its transmission is approximated to a Lorentzian function:

$$h(v) = T_0 / \left[1+(v)^2/(\Delta v_{FPI}/2)^2\right], \qquad (3)$$

where $\Delta v_{FPI}$ is the full width at half maximum of the transfer function, $T_0$ is the maximum transmission factor given by

$$T_0 = a_t(1-r_f)^2 / (1-a_t \cdot r_f)^2, \tag{4}$$

Where $a_t$ is the attenuation factor of the light intensity between travelling from one plate to the other, $r_f$ is the intensity reflection coefficient of the plates. Since the Brownian motion of aerosol particles does not broaden the spectrum significantly, the spectrum of aerosol backscatter $I(\upsilon,\upsilon_c,\upsilon_D,\Delta\upsilon_M)$ has nearly the same shape as the spectrum of the outgoing laser. Thus, the spectrum of aerosol backscatter can be approximated by a Gaussian function:

$$I(\upsilon,\upsilon_c,\upsilon_D,\Delta\upsilon_M) = (\sqrt{\pi}\Delta\upsilon_M)^{-1}\exp\left[-(\upsilon-\upsilon_c-\upsilon_D)^2/\Delta\upsilon_M^2\right], \tag{5}$$

Similarly, the measured reflection spectrum $R(\upsilon,\upsilon_c,\upsilon_D,\Delta\upsilon_M)$ can be expressed as follow:

$$R(\upsilon,\upsilon_c,\upsilon_D,\Delta\upsilon_M) = r(\upsilon) * I(\upsilon,\upsilon_c,\upsilon_D,\Delta\upsilon_M), \tag{6}$$

where $r(\upsilon) = 1 - h(\upsilon)$ is the reflection function of the FFP-SI. Then, a frequency response function Q is normalized and defined as

$$Q(\upsilon,\upsilon_c,\upsilon_D,\Delta\upsilon_M) = \frac{a^* \cdot T(\upsilon,\upsilon_c,\upsilon_D,\Delta\upsilon_M) - R(\upsilon,\upsilon_c,\upsilon_D,\Delta\upsilon_M)}{a^* \cdot T(\upsilon,\upsilon_c,\upsilon_D,\Delta\upsilon_M) + R(\upsilon,\upsilon_c,\upsilon_D,\Delta\upsilon_M)}, \tag{7}$$

where $a^*$ is a system constant, which can be determined in the calibration. By applying a least squares fit procedure to the Q function, the center frequency $\upsilon_c$ and the bandwidth $\Delta\upsilon_M$ can be retrieved simultaneously. The performance of the FFP-SI is characterized by spectrally analyzing the result from the Fig. 1(a). Since the linewidth of the monochromatic CW laser (3 kHz) is much less than that of the $\Delta\upsilon_{FPI}$ (~94 MHz), the measured spectrum represents the feature of the FFP-SI. Using the measured data near the first resonance peak (as shown in Fig. 1(b)), the frequency response function Q can be calculated by using Eq. (7). The calculated Q function and fitting result are presented in Fig. 1(c). The fitted $\Delta\upsilon_{FPI}$ is 94 MHz. In this calibration experiment, the overall transmission efficiency and reflection efficiency of the FFP-SI are measured to be 30 % and 58 % respectively.

    The HSRWL has several advantages over the coherent detection lidar as mentioned in introduction. However, the HSRWL has also several technical challenges, which are not present in its coherent counterpart. For weak backscattered signal from the aerosol, detection with high signal-to-noise ratio is required. To address this challenge, high efficiency and low noise upconversion detectors are adopted, which convert the working wavelength from 1548 nm to 863 nm and then detect with Si-APDs. In addition, the frequency stability control of the direct detection Doppler wind lidar system is a challenging task [22, 30]. Here, the frequency

drift of the FFP-SI relative to the outgoing laser is controlled by putting the interferometer inside a chamber with temperature stability of ±0.001 °C during each scanning process.

## 3. Instrument

Schematic diagram of the HSRWL is shown in Fig. 2. A continuous wave from a seed laser (Keyopsys, PEFL-EOLA) is chopped into pulse train after passing through an electro-optic modulator (EOM) (Photline, MXER-LN, extinction ratio 40 dB). The EOM is driven by a pulse generator (PG), which determines the shape and pulse repetition rate (12 kHz) of the laser pulse. A small fraction of the pulse laser is split out as a reference signal by using a polarization-maintaining fiber splitter. The reference signal is attenuated to the singe-photon level by using a variable attenuator ($VA_0$) and then fed to the receiver. The main laser pulse from the other output of the splitter is fed to an erbium-doped fiber amplifier (Keyopsys, PEFA-EOLA), which delivers pulse train with pulse energy of 50 μJ and pulse duration of 200 ns. A large mode area fiber is used to increase the threshold of stimulated Brillouin scattering (SBS) and self-saturation of amplified spontaneous emission (ASE). The pulsed laser beam is collimated and transmitted to the atmosphere by an 80 mm diameter off-axis telescope. The outgoing laser and atmospheric backscatter are separated by an optical switch composed of a quarter-wave plate and a polarizing beam splitter (PBS).

An optical circulator is used to separate the transmitted signal and the reflected signal of the FFP-SI. The FFP-SI is made of single-mode fiber, thus two polarization controllers are added at the front and rear ends of the FFP-SI to eliminate the polarization dependent loss. The continuous wave from the pump laser at 1950 nm is followed by a thulium-doped fiber amplifier (TDFA), both manufactured by AdValue Photonics (Tucson, AZ). The residual ASE noise is suppressed by using a 1.55 μm/1.95μm wavelength division multiplexer ($WDM_1$). The pump laser is split into two beams by a 3 dB fiber splitter, one used for the transmitted signal and the other for the reflected signal. The reflected backscatter signal and the pump laser are coupled into a periodically poled Lithium niobate waveguide ($PPLN-W_1$) via the $WDM_1$. Similarly, the transmitted backscatter signal and the pump laser are coupled into a $PPLN-W_2$ via the $WDM_2$. Optimized quasi-phase matching condition is achieved by tuning the temperature of the PPLN-W with a thermoelectric cooler [31]. Here, the upconversion detectors are integrated into an all-fiber module, in which the PPLN-Ws are coupled into polarization-maintaining fiber/multimode fiber (MMF) at the front/rear end. The backscatter signal photons at 1548 nm are converted into sum-frequency photons at 863 nm and then picked out from the pump and spurious noise by using an interferometer filter (IF) at 863 nm with 1nm bandwidth. Finally, the photons at 863 nm are detected by using Si-APDs (EXCELITAS). The TTL signals corresponding to the received photons are recorded on a multiscaler (FAST ComTec, MCS6A) and then transferred to a computer. The system detection efficiency of the upconversion detector used to detect the transmitted signal is tuned to 20.5 % with a noise level of 300 Hz. For another upconversion detector, the system detection efficiency is tuned to 15 % with a noise level of 330 Hz.

It is useful to understand how the different signals are detected through a data acquisition timing sequence, as shown in Fig. 3. A synchronization transistor-transistor logic (TTL) output from the PG is applied to synchronize the range-gating electronics. The signal collected in ten bins before the laser pulse is used to measure the averaging background. The reference signal from the $VA_0$ is collected in the three bins. The output of the Si-APD is disabled when a low level of TTL is applied to its gate input. Thence the trigger to the Si-APDs can be used to eliminate the strong mirror reflections from the telescope. The atmospheric backscatter signal is delayed by a PMF of 160 m to avoid mixing with the reference signal.

It is indispensable to control the temperature of the FFP-SI precisely, due to its temperature-sensitive characteristics. In this work, the FFP-SI is cased in a chamber, and two-stage temperature controller is applied. Temperature of the chamber is controlled by using Peltier elements with a PID temperature controller (BelektroniG B-20).

In order to investigate the effect of the temperature drift on measurement, the inner controller is shut down, and the frequency response function is measured by scanning the cavity length with the reference laser pulse over 95 seconds. By applying a least-square fit procedure to the measured frequency response functions, the center frequencies and bandwidths are retrieved simultaneously. As shown in Fig. 4(a), the center frequencies (circles) and temperature drift (line) are monitored over 6 hours.

Obviously, the temperature of the FFP-SI will fluctuate obviously as show in Fig. 4, even the ambient temperature inside the lab is $25 \pm 1\,°C$. Also the measured bandwidth is compared with the temperature drift ($\Delta T$) at a single scanning process, as plotted in Fig. 4(b). One can noted that large temperature drift will seriously affect the accuracy of the bandwidth measurement. Fortunately, the standard deviation of the measured bandwidth is 0.6 MHz when the temperature drift stabilize within $0.004\,°C$. When the inner PID control of the temperature is turn on, a stability of $\pm\,0.001\,°C$ is realized in ten minutes, as shown in Fig. 5. Consequently, the system error induced by the temperature fluctuation can be ignorable when the temperature controller is performed.

## 4. Field Experiments

Wind detection is carried out at Hefei (31.83 °N, 117.25 °E) in Anhui province, China. The location is 29.8 m above the sea level.

In order to obtain the aerosol backscatter spectra, the FFP-SI is scanned over 39.30 nm /297 MHz with a step of 1.31 nm / 9.9 MHz, by changing the driven voltage. At each step, backscatter signals of 22000 laser pulses are accumulated. Taking the time for data processing into account, each step costs 2 seconds. Considering 30 steps, the sampling of the entire aerosol backscatter spectra takes 1 minute. An example is shown in Fig. 6, both the transmitted spectra and the reflected spectra of the backscatter at different distance are recorded. Then the corresponding Q functions are calculated. An example of the calculated Q functions of backscatter at 0.03 km and 1.80 km are shown in Fig. 7(a), the corresponding fitting results are plotted with lines. The fitting residuals are plotted in Fig. 7(b). Profiles of transmitted and reflected backscatter along the distance indicate the quality of the measured data. As shown in Fig. 7(c), during the scanning, the highest reflected backscatter profile $N_R(\upsilon_1)$ and the weakest transmitted backscatter profile $N_T(\upsilon_1)$ at the beginning step of the scanning are plotted. Also the highest transmitted backscatter profile $N_T(\upsilon_2)$ and the weakest reflected backscatter profile $N_R(\upsilon_2)$ relative to the peak position of the Q function of the reference signal are present. It is obvious that detection of both the reflected and transmitted signal make full use of the backscatter. Judging from the raw signal, one can see that the detection range of 4 km can be reached.

Continuous LOS wind speed observations from the HSRWL are performed from 15:00 PM to 21:00 PM on Apr. 13. 2016 and on Apr. 14. 2016. As indicated in Fig. 8, $\theta$ is the angle of direction from the north in clockwise. The telescope points at a fixed azimuth angle of -75° on Apr. 13. 2016, and it points at 70° on Apr. 14. 2016. The experiment results are shown in Fig. 9. The distance resolution is switched from 30 m to 60 m at a distance of 1.8 km. The time resolution is 1 minute. One interesting phenomenon is the sign of the wind speed changed at sunset on Apr. 13. 2016. However, the LOS wind speeds show irregular distribution on Apr. 14. 2016. The different observation phenomena between the two consecutive days are relative to the telescope pointing and wind direction. On the first day, the detected region is flat, and the laser beam passed on the top of buildings. Therefore, the detection result reflects the actual state of the atmospheric wind field. However, on the second day, the laser beam passes through the gaps among the buildings, thus the wind field is disrupted.

To estimate the accuracy of the measurements, radial wind velocities from an ultrasonic wind sensor (Vaisala windcap WMT52) are compared with the lidar radial velocity measurements in the first bin, as shown in Fig. 10. Two sets of LOS wind speed measurements carried out on Apr. 13 and Apr. 14. 2016 show good agreement.

For quantitative analysis of the correlation of the two sets of the measurements, the scatterplot of the wind components from the HSRWL and the Vaisala is given in Fig. 11(a). The R-square and slope of the linear fit are 0.996 and 0.991 respectively. Fig. 11(b) is the histogram distribution of the wind difference between the results from two instruments. The mean difference and the standard deviation of LOS wind speed are 0.01m/s and 0.50 m/s respectively.

One of the attractive aspects of the proposed lidar is that not only the Doppler shift but also the bandwidth of the aerosol spectrum can be retrieved simultaneously. Typical profiles of the LOS wind speed and the bandwidth variation are plotted in Fig. 12(a) and Fig. 12(c) respectively.

As shown in Eq. (2) and Eq. (6), $T(v)$ function and $R(v)$ function are the Voigt functions, as a result of the convolution of a Gaussian function and a Lorentzian function. Due to the computational expensive of the convolution operation, it is difficult to give an analytic expression of the errors of the wind speed and the bandwidth variation. Thus, an empirical estimation is adopted. As an example, 30 continuous wind profiles and bandwidth variation from 18:00 to 18:30 on Apr. 13. 2016 are selected, since the wind field is relatively stable during the period, as shown in Fig. 9. Then a subtraction of two adjacent profiles is performed, and the results are plotted as circles in Fig. 12(b) and Fig. 12(d). The means of the difference at different distance are plotted as yellow line, and the standard deviations of the subtraction ($\sigma_-$) are plotted as red line in Fig. 12(b) and Fig. 12(d).

In probability theory, if X and Y are independent random variables that are normally distributed, then their difference is also normally distributed. And the square of the standard deviation of their difference is the sum of the square of the two standard deviations, i.e.

$$\sigma^2(X-Y) = \sigma^2(X) + \sigma^2(Y). \tag{8}$$

Assuming the adjacent measurements are independent, then the standard deviation of the measurements $\sigma_m$ can be described as:

$$\sigma_m = \sigma_- / \sqrt{2}. \tag{9}$$

In Fig. (12), one can see that, the standard deviation of the wind speed is 0.76 m/s at 1.8 km. And the standard deviation of the retrieved bandwidth variation is 2.07 MHz at 1.8 km.

To show the difference of the bandwidth variation between two days, the bandwidth variations from 0 m to 480 m are analyzed statistically. Specifically, Fig. 13(a) is the histogram distribution of the bandwidth variation on Apr. 13. 2016, and Fig. 13(b) is the histogram distribution of the bandwidth variation on Apr. 14. 2016. The mean and standard deviation of the bandwidth variation are 2.73 MHz and 2.55 MHz on Apr. 13. 2016. The mean and standard deviation of the bandwidth variation are 3.03 MHz and 2.94 MHz on Apr. 14.

2016. The bandwidth variation is directly related to the atmospheric turbulence, and can be used to estimate turbulence energy dissipation rate (TEDR) [7], which is beyond the scope of this article.

## 5. Conclusion and future research

An all-fiber, eye-safe and micro-pulse direct detection Doppler lidar has been demonstrated by using a high spectral resolution FFP-SI and two high signal-to-noise upconversion detectors. In experiments, an active temperature controller was adopted, which ensures temperature stability within $\pm 0.001\ °C$ during the scanning. Then the effect of thermal fluctuations on the FFP-SI was ignorable. Through spectral analyses of the backscatter signal and reference laser pulse, the Doppler shift and bandwidth variation was retrieved simultaneously. Continuous observation of LOS wind speed was performed. In the comparison experiments, the LOS wind measurements from the HSRWL and the Vaisala were in good agreements. Statistics analysis of the FHWM variation was carried out when the telescope pointed to different directions. In the future research, a matching azimuth-over-elevation scanner will be installed to realize full sky access. And the bandwidth variation will be used to study the atmospheric turbulence. In the meantime, the high temporal and spatial resolution wind data shows great potential in many applications, including measurement of aircraft wake vortices, wind shear, microbursts and true airspeed.

## Acknowledgements

This work was supported by National Fundamental Research Program (2011CB921300, 2013CB336800); National Natural Science Foundation (41274151, 41421063); CAS Hundred Talents Program (D); CAS Program (KZZD-EW-01-1) and the 10000-Plan of Shandong Province; the Fundamental Research Funds for the Central Universities (WK6030000043).

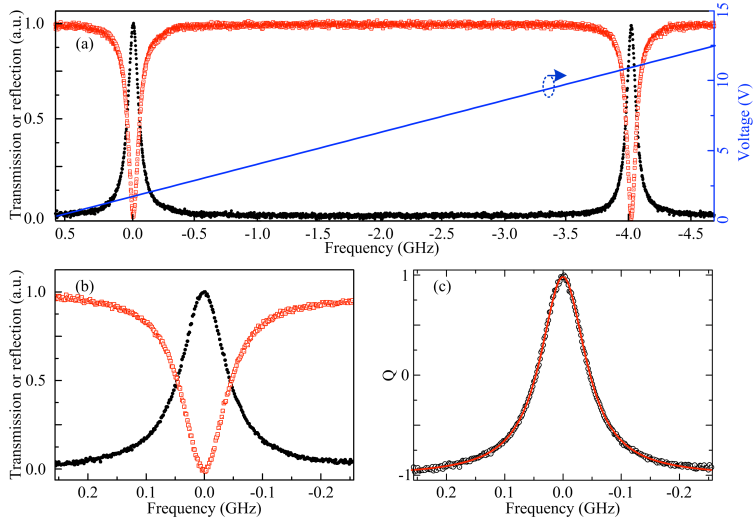

Fig. 1. (a): Transmission spectrum (circles) and reflection spectrum (squares) of the monochromatic CW laser through the FFP-SI as the driven voltage increases from 0.3 V to 12.5 V, (b): Zoom-in image of Fig. 1(a), (c): The calculated Q function (circles) and its Lorentzian fitting result (solid line).

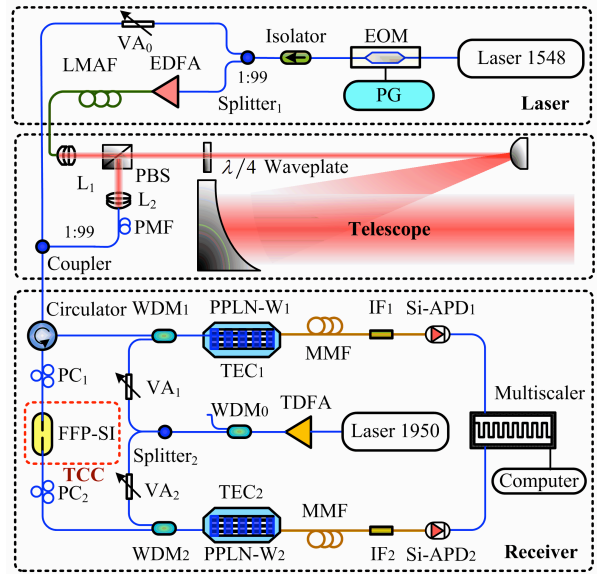

Fig. 2. System layout. EOM, electro-optic modulator; PG, pulse generator; VA, variable attenuator; EDFA, erbium doped fiber amplifier; LMAF, large-mode-area fiber; PBS, polarizing beam splitter; PMF, polarization-maintaining fiber; PC, polarization controller; TCC, temperature controlled chamber; FFP-SI, fiber Fabry-Perot scanning interferometer; TDFA, thulium doped fiber amplifier; WDM, wavelength division multiplexer; PPLN-W, periodically poled Lithium niobate waveguide, MMF, multimode fiber; TEC, thermoelectric cooler; IF, interferometer filter.

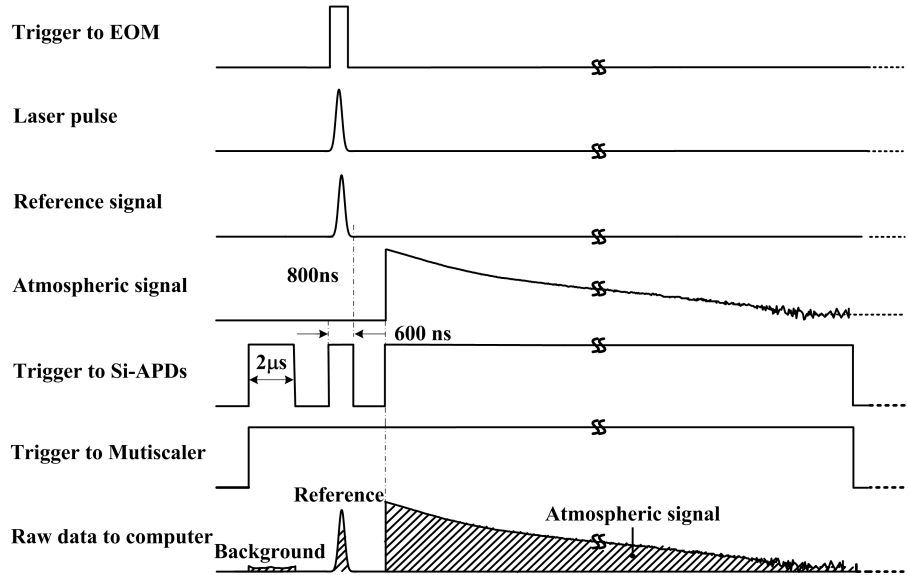

Fig. 3. Data acquisition timing sequence for a single laser pulse

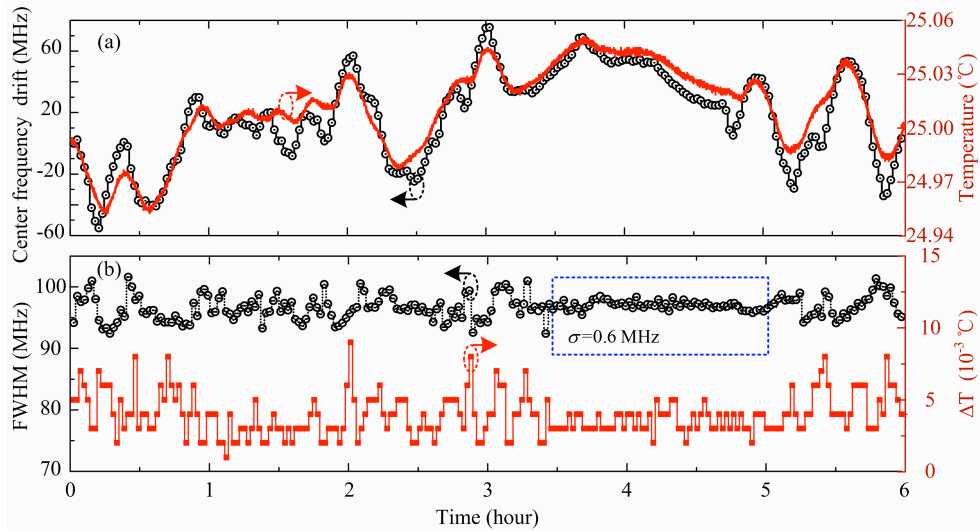

Fig. 4. Frequency drift (a) and bandwidth change (b) of the FFP-SI due to temperature fluctuation.

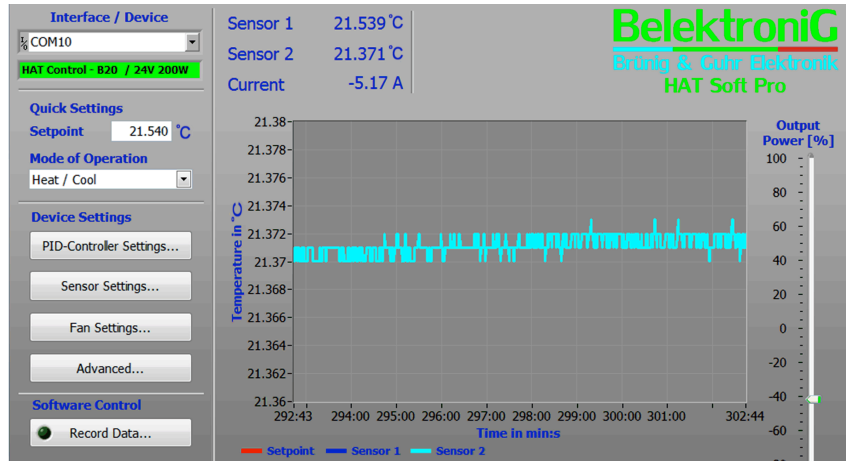

Fig. 5. Graphical user interface of the temperature controller (temperature stability of ±0.001 °C in ten minutes is realized)

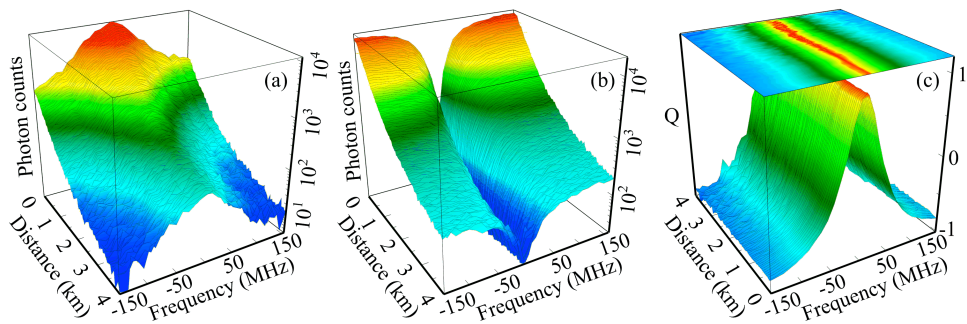

Fig. 6. Photon counts of transmitted backscatter signal (a), reflected backscatter signal (b) and the corresponding frequency response functions (c) along the 4 km.

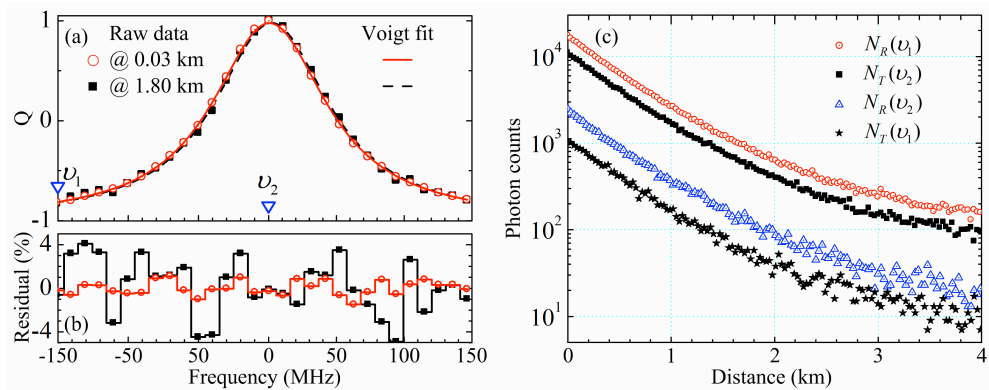

Fig. 7. (a): Frequency response functions of the backscatter aerosol at 0.03 km (open circle) and 1.80 km (filled squares) and their best fit results (solid line and dashed line), (b): Residual between the calculated frequency response functions and its fit results, (c): Profiles of transmitted and reflected backscatter along distance at given frequencies labeled in Fig. 7(a).

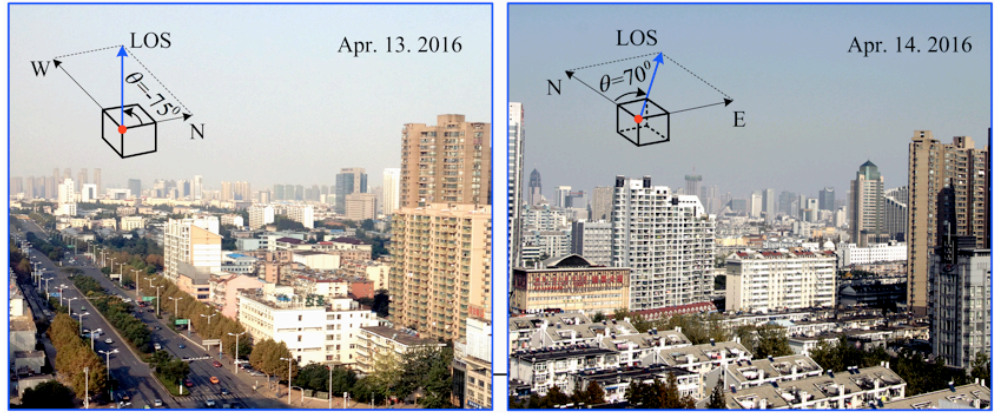

Fig. 8. Photos indicate the direction of the laser beam

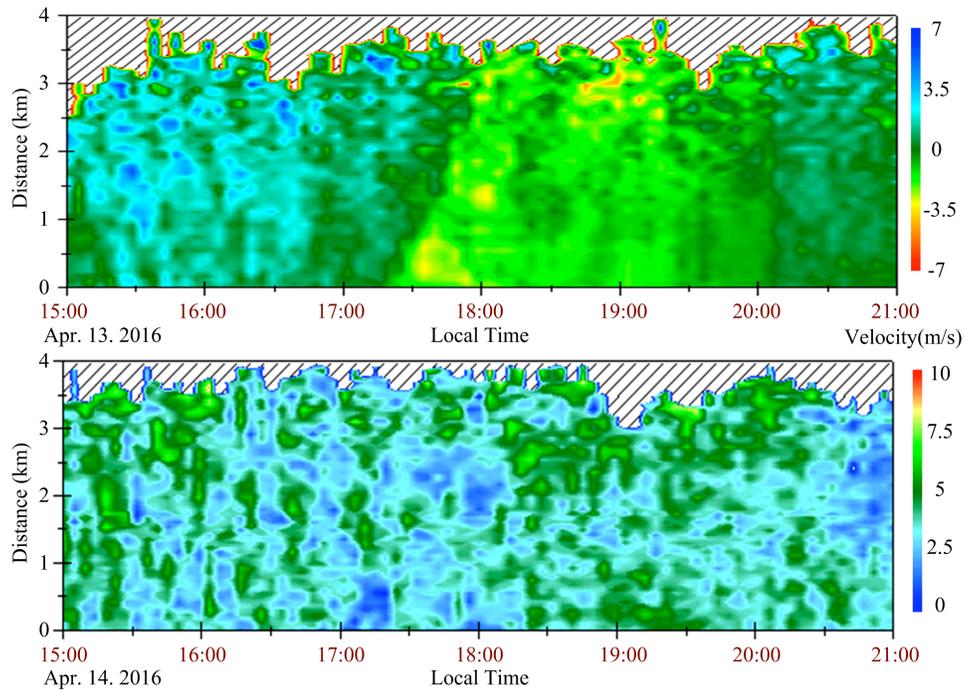

Fig. 9. Time-distance plots of continuous observation of the LOS wind speed from the HSRWL on Apr. 13. 2016 (top) and on Apr. 14. 2016 (bottom).

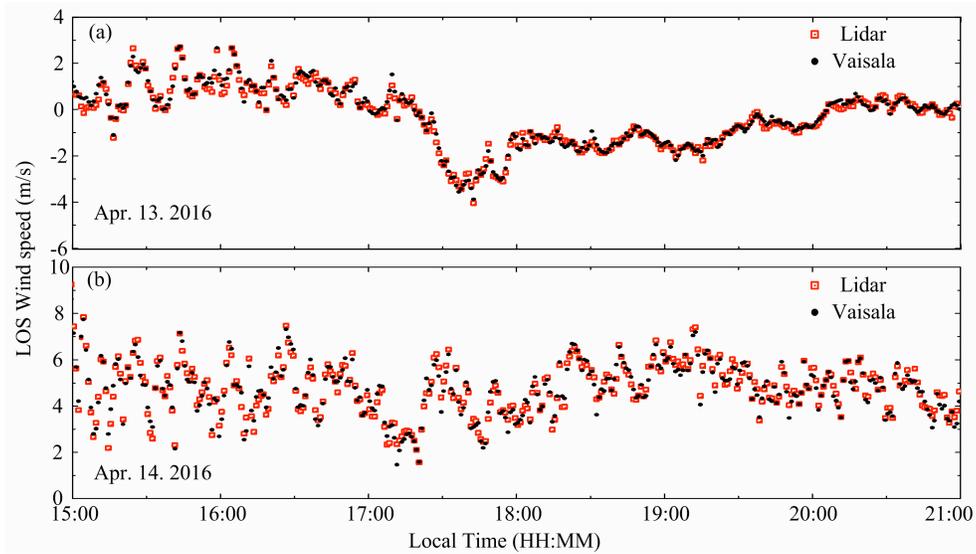

Fig. 10. LOS wind speed measured by the HSRWL (open squares) compared with data measured by the Vaisala (filled circle.) on Apr. 13. 2016 (top) and on Apr. 14. 2016 (bottom).

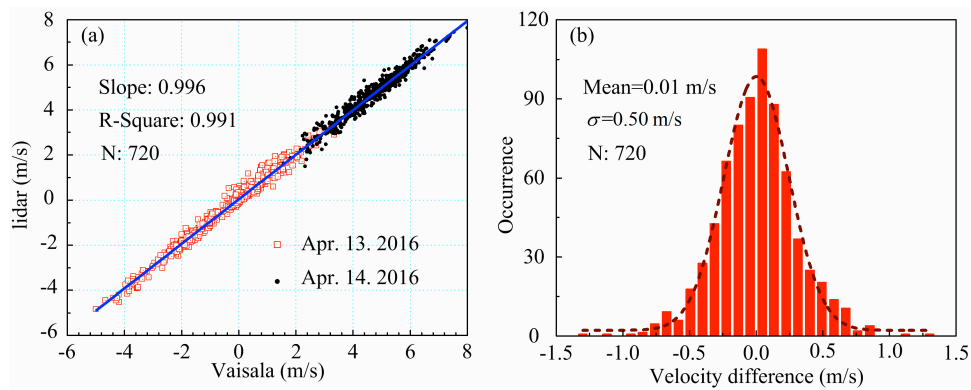

Fig. 11. (a): Scatterplot of wind speed from the HSRWL and Vaisala, (b): Histogram distributions of the velocity difference between the HSRWL measurements and the Vaisala measurements (red line is the Gaussian fit result to the data).

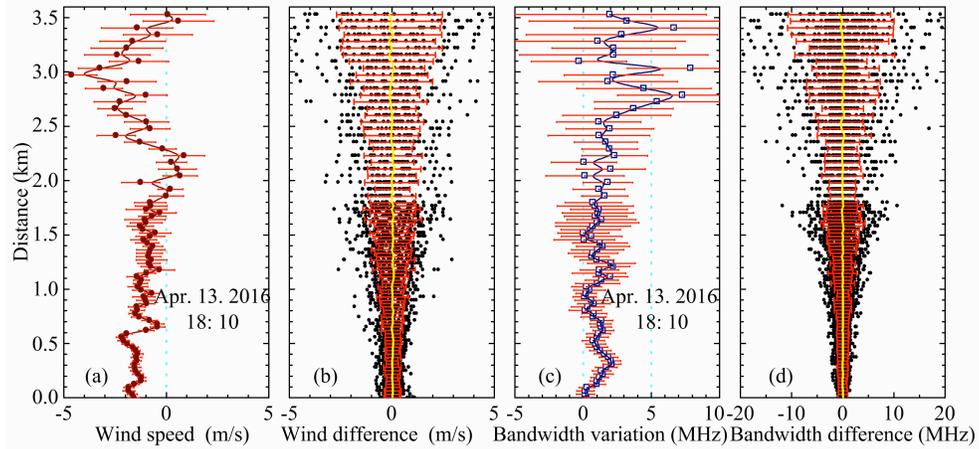

Fig. 12. Profiles of the LOS wind speed (a) and bandwidth variation (c), and the corresponding estimated measurement error (b and d) by using an experimental method.

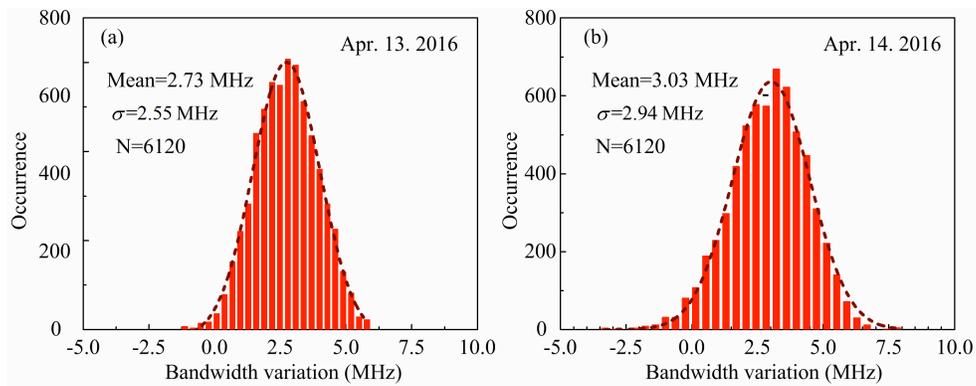

Fig. 13. Statistics of the bandwidth variation: histogram distributions of bandwidth variation from 0 m to 480 m on Apr. 13. 2016(a) and on Apr. 14. 2016(b), solid lines are the Gaussian fit results to the data.